# ARTICLE

# pyCOFBuilder: A python package for automated creation of Covalent Organic Framework models based on the reticular approach

Felipe Lopes Oliveira [a] and Pierre Mothé Esteves [a,*]

* Corresponding author: pesteves@iq.ufrj.br

Covalent Organic Frameworks (COFs) have gained significant popularity in recent years due to their unique ability to provide a high surface area and customizable pore geometry and chemistry. These traits make COFs a highly promising choice for a range of applications. However, with their vast potential structures, exploring COFs experimentally can be challenging and time-consuming, yet it remains an attractive avenue for computational high-throughput studies. However, generating COF structures can be a time-consuming and challenging task. To address this challenge, here we introduce the pyCOFBuilder, an open-source Python package designed to facilitate the generation of COF structures for computational studies. The pyCOFBuilder software provides an easy-to-use set of functionalities to generate COF structures following the reticular approach. In this paper, we describe the implementation, main features, and capabilities of the pyCOFBuilder demonstrating its utility for generating COF structures with varying topologies and chemical properties. pyCOFBuilder is freely available on GitHub at https://github.com/lipelopesoliveira/pyCOFBuilder.

## 1. Introduction

The quest for understanding the properties and discovery of new materials has been the subject of intense interest to the scientific community for many years. Although most discoveries have been made through trial and error combined with serendipity, the design of materials at the atomic level and the control over their properties at the nanometric scale has been the quintessential objective of materials scientists.

A class of materials that have attracted significant interest over the past years is the Covalent Organic Frameworks (COFs). COFs are materials with well-defined nanoporous architectures designed in a bottom-up approach by the covalent bonding of one or more organic building blocks by strong covalent bonds, thus forming an extended nanoporous crystalline structure.[1–3]

The general process for generating a structural model for a COF structure is commonly referred to as the "reticular approach", as depicted on Fig. 1.[4–7] This method starts by selecting a suitable network topology and breaking it down into its building units. The geometric constraints of the underlying network determine the building block geometry (linear, triangular, squared, etc.) and connectivity (number of points used to build the extended structure). The building blocks are created based on an organic core that defines its size and chemical properties. Next, a suitable set of reacting groups is selected to create the covalent connections between the core units to form the extended reticular structure. Finally, functional groups can be introduced into the organic core, either *pre-* or *pos-* synthetically, to control the physical-chemical characteristics of the pore surface.

The use of the reticular approach to construct crystalline organic structures provides five different dimensions for the design of materials, i.e. topology, organic core, connection group, functionalization, and supramolecular arrangement, thus allowing the control at the atomic level over the material properties. This makes the COFs attractive for a wide range of applications that include gas capture, separation, and storage,[8–10] heterogeneous catalysis[11–13], energy storage and production,[14,15] chemo-sensing,[16] organic semiconductors,[17] supercapacitors,[18] and many others.[19,20]

Currently, a considerable amount of different COFs have been successfully synthesized and characterized, with these structures compiled in several databases.[21,22] However, the number of reported structures barely scratches the surface of the chemical diversity that could be generated. The combination of structural diversity presented by different topologies with the extensive library of organic molecules available for the construction of COFs combined with the additional variability introduced by different connection groups and functionalizations makes the chemical space of available structures formidably large.[6,23]

Much of the research done until recently is profoundly based on the trial-and-error approach, where a combination of basic knowledge and chemical intuition is applied to propose new materials or modifications to existing ones, with a great research effort dedicated to synthesizing and characterizing the desired structures. However, due to the high financial cost, enormous human effort, and time required for this strategy there has been a growing interest in the development of alternative methodologies for materials discovery.[6,24,25]

Recent advancements in computational chemistry, coupled with the increasing processing power of modern computational facilities, have enabled the use of high-throughput computation-based and machine-learning approaches.[26–28] These approaches require the analysis of a vast number of chemical and structural variations for numerous compounds, making it possible to select only the most promising candidates for further synthetic efforts. This powerful combination of computational tools and the ability to understand at a molecular level the phenomena related to the applications of COFs helps to streamline the materials discovery process, allowing researchers to more efficiently explore, identify, and improve new useful materials. However, generating COF structures for computational studies can be a challenging and time-consuming task, requiring a high level of expertise in both molecular modeling and computer programming.

Currently, the options for creating computational models for reticular materials are notably constrained, being limited to codes developed to generate specifically Metal-Organic Frame-work structures or focused on the generation of COFs from a specific set of building blocks.[29–34] In 2018, Mercado et al.[35] developed a database containing 69,840 structures created from 666 distinct building blocks and utilized GCMC calculations to assess the most suitable candidates for methane capture. In 2018, Lan et al.[36] introduced a database comprising approximately 470,000 materials. In 2023, De Vos et al.[37] presented a database named ReDD-COFFEE, which encompassed 268,687 COF structures along with system-specific force fields. Although, in the latter case, scripts for creating the structures present in the database have been made available, the lack of a systematic nomenclature for the creation of structures and for the representation of chemical features such as functional groups or connector groups can hinder their use for generating new structures not present in the original database.

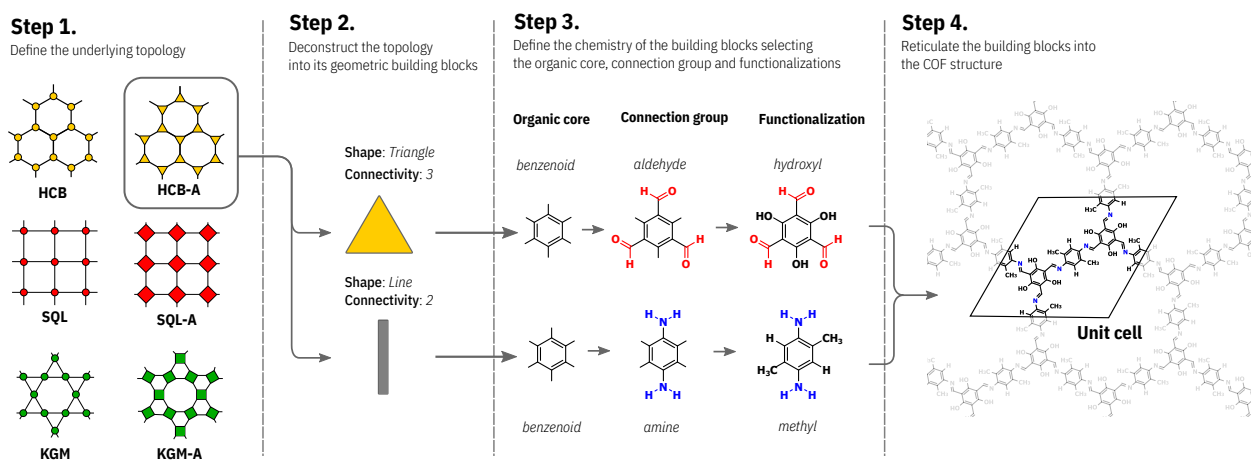

**Figure 1: Reticular approach for building a COF material.** This process involves **1.** selecting an underlying topology and a net that represents it, **2.** decomposing a given net into nodes and linker, **3.** selecting connecting chemistry and functionalization to form the building block molecule, and **4.** create covalent bonds between these building block molecules (reticulation) according to the geometry dictated by the net to form an extended structure represented by a unit cell.

Here, we introduce pyCOFBuilder an open-source Python package that automates the creation of computational models for COF structures based on the reticular approach. The current version of pyCOFBuilder offers a user-friendly platform with a broad range of features, including the implementation of main 2D and 3D networks, various stacking patterns (for 2D structures) or interpenetration classes (for 3D structures), a vast library of building blocks with dozens of possible functionalizations. With these features combined, pyCOFBuilder can potentially generate billions of unique COF structures, thus offering a robust path for implementing diverse computational and machine-learning techniques in developing new COF materials.

To illustrate the power of the reticular approach unlocked by the use of pyCOFBuilder, we generated a diverse set of structures exploring the four dimensions of the reticular approach and investigated their impact on the $CO_2$ capture capability of these materials. The results demonstrate that it is not only possible to generate materials with higher capture capacity than those currently synthesized but also to utilize machine learning techniques to dissect the contribution of each dimension of the reticular design in this property, thereby enabling a deeper understanding of the $CO_2$ capture phenomenon by COFs.

## 2. Computational Methods

To evaluate the quality of the structures generated by pyCOFBuilder the obtained structures were fully optimized at the DFT PBE-D3(BJ) level and using two different tight-binding approaches: the density functional tight binding (DFTB)[38] with mio set of parameters[39] and the extended Tight Binding (xTB)[40] with the GFN1-xTB method, both implemented on CP2K.

The DFT calculations were performed using the Perdew–Burke–Ernzerhof (PBE) exchange-correlation functional[41] with DFT-D3(BJ) dispersion corrections.[42,43] The Quickstep code of the CP2K v2023.1 package was used,[44,45] employing GTH pseudopotentials,[46,47] TZV2P-MOLOPT contracted Gaussian basis sets, and an auxiliary plane wave basis set. The plane wave cutoff is set to 600 Ry cutoff mapped on a 5-level multigrid. The orbital transformation (OT)[48] method was used for all the structures with band gap > 0.5 eV. For the structures with a band gap < 0.5 eV, the Broyden diagonalization and Thomas–Fermi smearing with 300 K for electronic temperature.

For the geometry optimization, the atomic positions and cell parameters were fully optimized until the convergence with total forces below 1.0 millihartree/bohr (a root-mean-squared value below 0.7) and the total pressure below 100 bar, using the Limited-memory BFGS (L-BFGS) minimizer. The unit cell was optimized, imposing the constraint that the original symmetry of the generated structure, i.e., the type of Bravais lattice, is maintained to allow a direct comparison with experimental data. Tests have demonstrated that in the majority of cases, the symmetry remains intact even without the need for this constraint during optimization. In the rare instances where there is some variation, it is typically small and does not affect any of the conclusions drawn.

With the obtained structure the textural properties, namely largest included sphere ($D_i$), largest free sphere ($D_f$), largest included sphere along free path ($D_{if}$), crystal density $\rho$, volumetric and gravimetric surface area and pore volume were calculated using the Zeo++ software v0.3 using a probe with radius of 1.86 Å.[49,50]

To simulate the gas uptake by the studied structures force field-based Grand-Canonical Monte Carlo (GCMC) simulations were performed using the RASPA2[51,52] package. For all GCMC simulations, 50,000 cycles were employed and the equilibrium uptake average was calculated only on the equilibrated part using the pyMSER package. The simulations were performed in a supercell large enough that each cell vector is larger than the cutoff radius for short-range interaction (12.8 Å). For the long-range interactions, the Ewald sum technique was used with a relative precision of $10^{-6}$. A Lennard-Jones potential with parameters taken from the TraPPE[53] force field was used to treat the van der Waals interactions of adsorbed molecules ($N_2$ and $CO_2$) and from DREIDING[54] to treat the framework atoms. LJ parameters between atoms of different types were calculated using the Lorentz-Berthelot mixing rules and were shifted to zero at the short-range cutoff.

The electrostatic interactions were calculated using partial charges centered in the atoms, calculated based on the Density-derived electrostatic charges (DDEC) approach within the DDEC6 method[55–58] as implemented in the Chargemol software. The charges were derived based on the electron density of the optimized structures computed by CP2K on the DFT PBE-D3(BJ) level.

## 3. Results and Discussion

### 3.1. pyCOFBuilder general algorithm.

**String-based representation for reticular structures** The primary aim of pyCOFBuilder is to provide a comprehensive solution for the generation of COF structures using the reticular approach. This approach involves the connection of organic building blocks by covalent bonds to create a crystalline structure with the geometry defined by an underlying net. The resulting structures can exhibit a variety of functional groups and intermolecular degrees of freedom, *i. e.*, distinct stacking pattern for 2D structures or interpenetrating class for 3D structures.

Due to the complexity of COF structures and the absence of a standardized nomenclature or string representation for crystalline structures, here we propose a simple yet effective string-based representation for COF structures. This representation is designed to encapsulate the key characteristics of a reticular structure, including the building blocks (`BB`), underlying net (`NET`), and stacking/interpenetration (`ST`) in the form of `BB1-BB2-NET-ST`.

**Table 1:** Symbols, connectivity numbers, and geometric figures used to represent the building blocks.

| Symbol | Connectivity | Geometric Figure |
|--------|--------------|------------------|
| L | 2 | Line |
| T | 3 | Triangle |
| S | 4 | Square |
| R | 4 | Rectangle |
| T | 4 | Tetrahedron |
| O | 6 | Octahedron |
| P | 6 | Trigonal prism |
| H | 6 | Hexagon |
| C | 8 | Cube |
| A | 8 | Square antiprism |
| E | 8 | Octagon |
| B | 12 | Cuboctahedron |
| I | 12 | Icosahedron |
| U | 12 | Truncated tetrahedron |
| X | 12 | Hexagonal prism |

The building blocks (`BB`) are represented by its four crucial properties: the geometry and connectivity number (`SC`), the type of organic core (`CORE`), the connection group (`CONECTOR`), and any present functional groups (`R_x`). This properties are represented by descriptors connected by underscores to form a unique string in the format `SC_CORE_CONECTOR_R_1_R_2_R_3....`. Here the underscores were chosen to make it easy to differentiate the descriptors of the structure, which is connected by dashes, from the descriptors of the building blocks. The symmetry and connectivity number (`SC`) are encoded by a letter that represents the geometry of the building block followed by its connectivity number, as shown in Table 1, using a modified version of the symbols used to represent vertex-transitive polyhedra.[59] Some examples of common organic cores, radical groups, and connection groups are presented in Fig. 2.

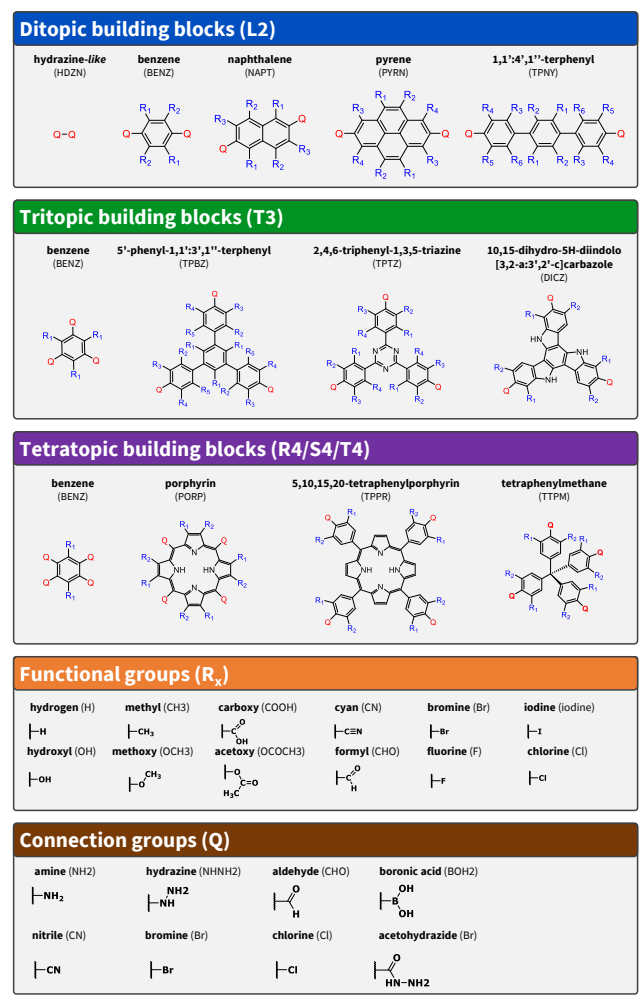

**Figure 2: Example of organic cores, connection groups, and functional groups available to generate COF structures.** The organic cores are encoded on the COF string representation using the four-letter code presented below the structures. Although arbitrarily defined, the letters are chosen to approximate an abbreviation of the IUPAC name of the molecule from which the organic core is derived. Q marks the position of the connection groups and $R_x$ marks the position of the functional groups. Both the connection group and functional groups are encoded using the common chemistry abbreviation.

To illustrate this representation, let us consider the 2,4,6-triformyl phloroglucinol molecule which is a commonly used building block for COF synthesis. In the proposed notation, this molecule is denoted as `T3_BENZ_CHO_OH`. Here, `T3` refers to the geometric shape and the number of connections of a tritopic building block, `BENZ` is a four-letter encoding of the benzenoid shape of the organic core, `CHO` represents the aldehyde groups that undergo a reaction with other building blocks to form the extended structure, and `OH` indicates the presence of hydroxyl groups in the 1, 3, and 5 positions.

To encode the reticular net a three-letter code adapted from the Reticular Chemistry Structure Resource (RCSR) nets is used.[60,61] The current version of pyCOFBuilder implements several different nets, as illustrated in Supporting Information Figure S2. To represent the stacking pattern or interpenetrating class desired, the encoding string will depend on the selected net. For 2D nets, available stacking patterns include `AA`, `AA1`, `AA2`, `AAl`, and `AAt`, among others deppending on the topology. In the case of 3D nets, the encoding string is determined by the number of interpenetrating structures.

The pyCOFBuilder nomenclature proposed here is, in a way,

```
1  import pycofbuilder as pcb
2
3  cof = pcb.Framework('T3_BENZ_CHO_OH-L2_BENZ_NH2_H-HCB_A-AA')
4  >> T3_BENZ_CHO_OH-L2_BENZ_NH2_H-HCB_A-AA                         hexagonal    P    P6/m  # 175    12 sym. op.
5  cof.save(fmt='cif', supercell=(1,1,2))
6
```

**Figure 3: Code snippet showing how to generate a cif file of a COF structure.** The Framework class creates the structure based on a given name. The save method saves the created structure with the desired file format and size of supercell.

inspired by the MOFid[62], RFcode[63], and the nomenclature proposed by Yaghi *et al.*[2] to represent MOFs and COFs. However, the pyCOFBuilder approach promotes a more simple alternative for the representation of COF structures, in addition to facilitating the incorporation of crucial information about the structure such as the differentiation of connection and radical groups and the information about the stacking/interpenetration profile of the final structure. This method can be readily expanded to represent Metal-Organic Frameworks using the secondary building unit (SBU) approach.[64]

The Supplementary Information provides the complete list of currently available organic cores, connection groups, functional groups, nets, and stacking/interpenetration options, along with examples for naming some well-known COFs.

**Structure building algorithm** After defining the string representation for the desired structure, pyCOFBuilder performs initial checks to ensure the construction is feasible. Specifically, it determines if the building blocks are compatible with the desired net and if the connection groups can be linked to each other. If the compatibility requirements are met, pyCOFBuilder proceeds to generate the crystalline structure.

The building block's structure for COFs is created by attaching the selected connection groups and functional groups to a chosen organic core, identified by a four-letter code. The organic core structures contain ghost atoms labeled as Q for connection groups and $R_x$ for functional groups. These ghost atoms act as positioning vectors directing the attachment of the selected groups on the organic core structure.

Each building block molecule is created as a BuildingBlock python object, that can be created and manipulated independently of the COF building process. The atomic positions of the building block structure can also be stored in $xyz$ format along with the structure during the creation of the COF, thus allowing independent calculations of these molecules independently of the COF structure.

Once the building blocks molecules are created, the selected net is used as a topological blueprint for the final structure. First, the size of the building blocks is used to calculate the cell vectors ($a$, $b$, and $c$). Next, the building blocks are rotated and translated to occupy the node and edge positions defined by the selected net. After the building blocks have been arranged in the correct positions, the pymatgen package is utilized to create a "Structure" object. This object is then subjected to symmetry analysis to calculate the space group and generate the crystal structure.

The stacking pattern can be specified using a two-letter code, such as AA, AA1, or AB1, while the number of interpenetrating structures for 3D nets is used to define the interpenetration class. To generate a COF structure with the desired stacking pattern or interpenetration class, the atoms in the unit cell are duplicated, rotated, and translated accordingly. Then, another symmetry analysis is performed on the structure to ensure that the correct unit cell is used to represent the new structure.

**Software outline and general usage** The pyCOFBuilder software was implemented using object-oriented Python and utilizes several scientific programming libraries, such as Numpy[65] and pymatgen[66] to execute mathematical operations and symmetry analyses. The GitHub repository contains detailed instructions about the manual and automatic installation. We recommend the user set up a Python environment using the environment.yml file provided therein.

The code is designed with a focus on two primary objects, namely BuildingBlock and Framework. The BuildingBlock object is intended to deal with the molecules that will form the COF and can be independently used. The Framework object represents the COF itself, which is constructed by linking the BuildingBlock molecules. The software is distributed under the MIT license and is provided with unit tests and comprehensive online documentation.

The COF structure, referred to in the code as Framework, can be created directly by their string-encoding as in the example shown on Fig. 3. For example, the COF referred to in the literature as TpPa-1[67] or DAB-TFP[68] can be translated to the unique representation T3_BENZ_CHO_OH-L2_BENZ_NH2_H-HCB_A-AA. The resulting COF structure generated by pyCOFBuilder can be saved in several file formats including CIF, XYZ, PDB, PQR, POSCAR, QuantumESPRESSO input, among others, allowing a simple integration on a wide variety of high-throughput workflows.

To assess the quality of the models generated by pyCOFBuilder, we selected a set of 33 structures from the literature and conducted a comparison of the experimental cell parameters with those generated by pyCOFBuilder, as well as those obtained after full geometry optimization (cell parameters and atomic positions) at DFT-PBE-D3(BJ) level and two Tight-Binding approaches (xTB-GNF1 and DFTB). The complete list of materials with their respective cell parameters and references are provided in *Supporting information*.

The results presented in Fig. 4 reveal that all tested methods exhibit good agreement with the experimental values for the lattice parameter a, showing root-mean-squared error (RMSE) values of 1.15, 1.01, 0.94, and 1.06 Å for pyCOFBuilder, DFTB, xTB, and PBE-D3(BJ), respectively. pyCOFBuilder generated structures with cell parameters close to the experimentally determined values even without undergoing any geometry optimization process, suggesting that the generated structures can be used effectively in high-throughput studies.

The crystalographic parameter c, on the other hand, exhibits a notable deviation from all the simulation methods applied to the experimental data. Whereas the experimental measurements exhibit values within the range between 3.1 and 4.1 Å, the simulated values tend to concentrate around certain specific values, which are 3.6, 3.1, 3.4, and 4.0 for pyCOFBuilder, DFTB, xTB, and PBE-D3(BJ), respectively. This discrepancy is attributed to the challenges in adequately treating the dispersion interactions between the COF sheets, which continues to pose a significant obstacle in accurately modeling 2D materials.[69,70] Thus, a more comprehensive analysis is required to better understand the underlying factors responsible for these variations and their implications on the properties obtained through different approaches.

**3.2. Property–Structure Relationships.** The application of the reticular approach to design new materials targeting spe-cific applications offers a significant advantage through its ability to exert atomic-level control over the structure of the material, which in turn enables precise manipulation of the material's properties. This control is achieved by exploring the five key dimensions of the reticular design: topology, organic core, connection group, functionalization, and supramolecular arrangement. By varying these dimensions, a wide range of materials with diverse chemical and geometrical attributes can be generated, ultimately leading to the development of materials with distinct and unique characteristics.

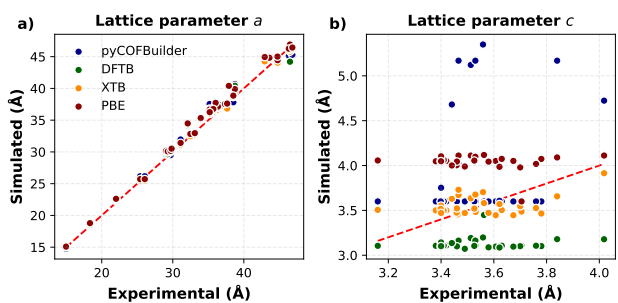

**Figure 4: Comparison between cell parameters of pyCOFBuilder generated structures, Tight-binding, and DFT-based geometry optimization with experimental data.** a) Lattice parameter a, b) Lattice parameter c. All materials possess hexagonal unit cells and thus $a = b$. It is noteworthy that the level of agreement between simulated and experimental values is substantially higher for parameter a compared to parameter c.

**Impact of the topology on textural properties** To exemplify the potential impact of the underlying topology on the textural characteristics of 2D COFs Fig. 5 presents the distribution of three crucial properties associated with the porosity of the material for six different 2D topologies. Notably, topologies such as KGD and HXL_A tend to decrease both pore size and pore volume, while concurrently increasing the specific surface area. This observation leads us to anticipate that materials possessing such distinct geometric arrangements might be particularly well-suited for applications demanding extensive surface areas, such as $H_2$ and $CH_4$ storage.

The augmented versions of the HCB and SQL nets present a clear tendency to increase both in pore size and specific area when compared to their non-augmented versions. Therefore, using these augmented networks can be a simple strategy to generate materials with larger pores.

**Impact of functional groups** To provide a concrete example of how pyCOFBuilder can expedite COF research, we investigate the potential use of several COFs for capturing $CO_2$. We selected two types of building blocks: tritopic blocks (BENZ and TBPZ) that contain aldehyde connector groups, and ditopic blocks (BENZ, NAPT, BPNY, DHSI, PYRN, ANTR, TPNY, DPEY, DPEL, DPBY, BPYB) that contain amine connector groups. Here, imine condensation was used to connect these building blocks as it is a widely used method for producing COFs with high crystallinity. The resulting COF structures presented an HCB-A topology. To introduce chemical diversity, we decorate the ditopic building blocks with 12 different functional groups: -H, -OH, -CH$_3$, -OMe, -OEt, -NH$_2$, -NO$_2$, -CN, -F, -COOH, -CHO, and -tBu.

The structures were built using pyCOFBuilder and optimized following the three-step procedure proposed by Ongari et al.[22] at the PBE-D3 level with CP2K. The partial charges for the atoms were calculated based on the electronic density of the optimized structure using the DDEC approach implemented on Chargemol. To evaluate the $CO_2$ adsorption capacity of the structures force-field-based Grand-Canonical Monte Carlo simulations were performed using the Dreiding force field to treat the framework atoms and the TraPPE force field for the adsorbed molecules. Simulations were performed at 298K with pressures ranging from 0.001 to 10 bar. For details on the simulation methodology see the Methods section. To evaluate the performance of porous materials for $CO_2$ capture applications the working capacity and the $CO_2/N_2$ selectivity were used as performance metrics.[71]

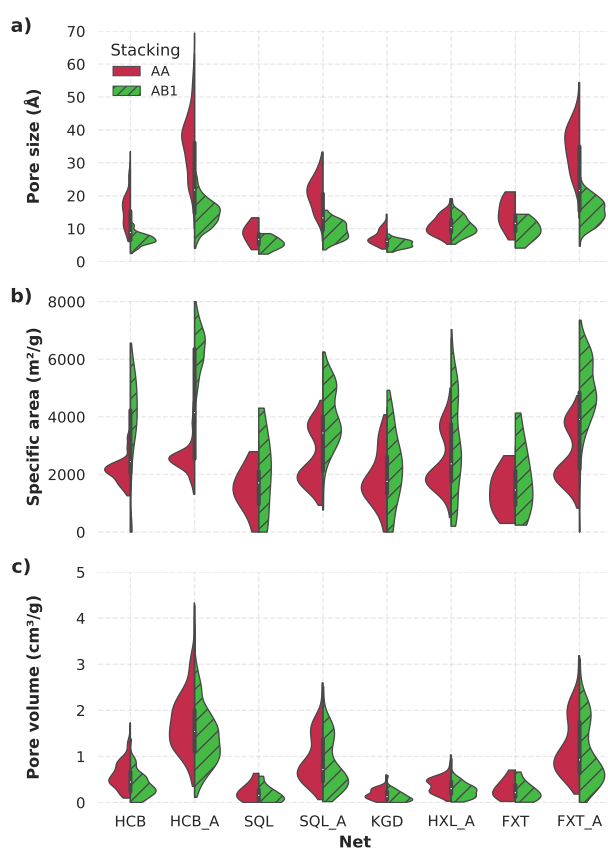

**Figure 5: Violin plot showing the influence of the topology and stacking pattern on three selected textual properties for 2D COFs.** a) Pore size, b) Specific area, c) Pore Volume.

The working capacity was calculated as the difference in the equilibrium amount adsorbed between the adsorption and desorption cycle design to model a simple PSA (*pressure swing adsorption*) cycle of $CO_2$ capture at 298K, with adsorption at 2 bar and desorption at 0.1 bar. The selectivity of $CO_2$ over $N_2$, $\alpha_{CO_2/N_2}$, is defined as

$$\alpha_{CO_2/N_2} = \frac{x_{CO_2}/x_{N_2}}{y_{CO_2}y_{N_2}} \quad [1]$$

where $x_i$ and $y_i$ are the adsorbed and gas phase mole fractions of species $i$, respectively

Fig. 7 shows the dependence of working capacity and selectivity on the textural properties of the selected COFs. Specifically, this dependence has been assessed concerning the COFs' specific area, pore volume, pore size, and void fraction. Although some trends are apparent regarding the metrics of performance and their dependence on these textural properties, the factor that appears to exert the most significant influence on the overall COF performance is the type of functional group. Indeed, by evaluating the isotherm of adsorption of some selected COFs (Fig. S8) it can be observed that the type of functional group is capable of producing variations on $CO_2$ uptake values ranging from 2.1 mol/kg (R = OEt) to 8.3 mol/kg (R=CHO) at 10 bar, despite the relatively small variation in pore size caused by the insertion of these functional groups. The impact of functional group choice is also reflected in the enthalpy of adsorption, which varies from 12 and 22 kJ/mol

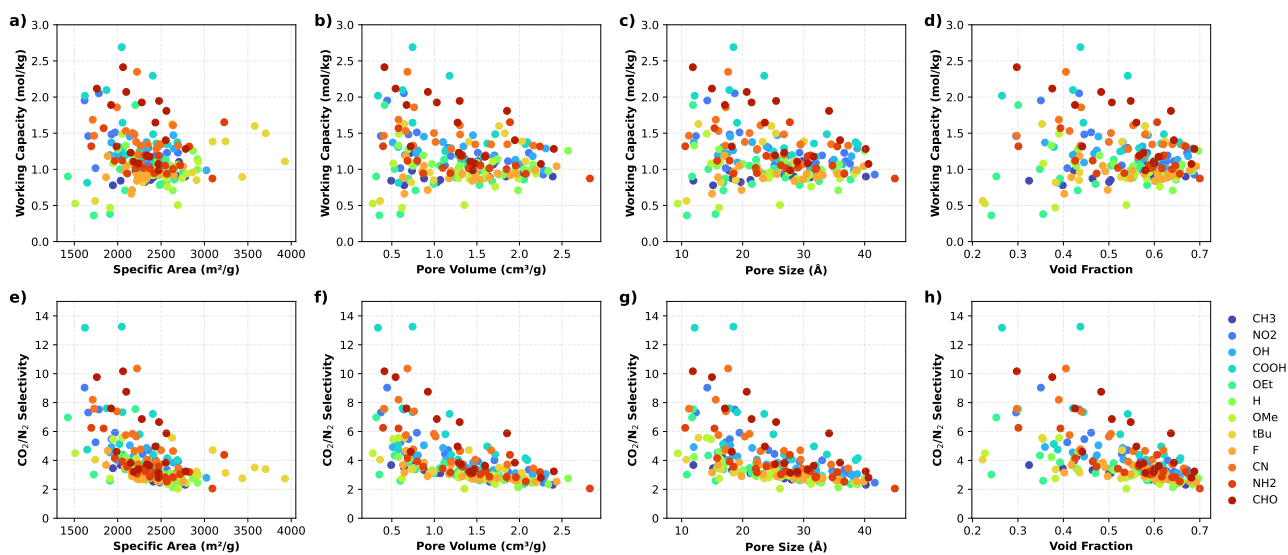

**Figure 6: Impatct of building block functionalization on $CO_2$ capture for different COFs.** The choice of the functional group presents a greater influence than the textural properties, such as specific area and pore volume, on the $CO_2$ working capacity (a-d) and $CO_2/N_2$ selectivity (e-h) for COFs.

for these functional groups.

To quantify the relative importance of the textural properties and functional group on the performance metrics, we combined the XGBoost machine learning algorithm with the SHAP (Shapley Additive exPlanations) approach. The XGBoost model was trained to predict the working capacity and $CO_2/N_2$ selectivity of a COF based on its textural properties and the type of functional group. The results, presented on Fig. 7, show that for the prediction of the working capacity, the functionalization has the higher global relative importance presenting a mean SHAP value of 0.16, followed by density, pore size, void fraction, specific area and pore volume with 0.14, 0.09, 0.09, 0.08, and 0.05 respectively. For the selectivity, the values followed the same trend with functionalization presenting a mean SHAP value of 0.61, followed by density, void fraction, pore size, specific area, and pore volume presenting values of 0.41, 0.38, 0.32, 0.23, and 0.18 respectively.

These results highlight the importance that the chemical characteristics present inside the pores can have in the design of new materials for $CO_2$ capture, indicating that this type of approach can be a viable path for developing new materials with high $CO_2$ capture capacity.

## 4. Conclusions

Here we have introduced pyCOFBuilder as a tool for the generation of COF structures for computational studies based on the reticular approach. This Python tool enables the generation of COF structures using a novel string-based representation proposed in this work.

The implementation details of the software have been presented in detail, including the object-oriented design and the scientific programming libraries utilized. Additionally, the generated structures demonstrated good agreement with experimental data and simulations with different levels of theory, highlighting the accuracy and reliability of the software.

We also presented an exploration of the property-structure relationships for the impact of the topology on the textural characteristics of 2D COFs and the working capacity for $CO_2$ capture and $CO_2/N_2$, two wildly used figures or merit for performance evaluation on carbon capture applications, highlighting the potential of pyCOFBuilder for practical use.

Overall, the presented results illustrate the utility of py-COFBuilder for the swift production of a wide array of COF

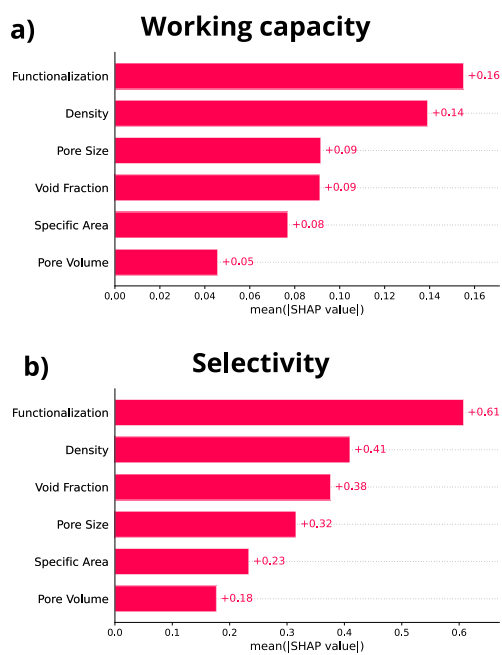

**Figure 7:** Global SHAP importance of each feature on the prediction of working capacity (a) and selectivity (b) by the trained XGBoost model. The functionalization type presents the higher means SHAP value for both properties, indicating its dominance over the textural properties.

structures. This process not only facilitates the identification of the most promising candidates for a particular application but also enables the scientific investigation of the underlying principles behind their performance.

We anticipate that by simplifying the process of generating COF structures, pyCOFBuilder can accelerate the discovery and optimization of new materials with desirable properties, ultimately facilitating progress and innovation in a diverse range of fields.

## Acknowledgments

We acknowledge financial support from CAPES (Project 001), CNPq, and FAPERJ. The authors would like to thank the Núcleo Avançado de Computação de Alto Desempenho (NACAD) of COPPE/UFRJ for the computational facility.

## Contributions

**Felipe Lopes Oliveira**: Conceptualization (lead); Visualization (lead); Writing – original draft (lead); Writing – review & editing (equal). **Pierre Mothé Esteves**: Conceptualization (supporting); Funding acquisition (lead); Supervision (lead); Writing – review & editing (equal).

All authors contributed to the discussion and approved the final version of the manuscript.

## Conflicts of interest

There are no conflicts of interest to declare.

## Data and Software Availability

The code presented in this manuscript is available under the MIT license at the GitHub repository https://github.com/lipelopesoliveira/pyCOFBuilder.

## Supporting Information

The Supporting Information is available free of charge on the publisher's website.

## Supplementary Information for
# pyCOFBuilder: A python package for automated creation of Covalent Organic Framework models based on the reticular approach

Felipe Lopes Oliveira  [a] and Pierre Mothé Esteves  [a]

*Corresponding author: pesteves@iq.ufrj.br*

[a]Instituto de Química, Universidade Federal do Rio de Janeiro, Av. Athos da Silveira Ramos, 149, CT A-622, Cid. Univ., Rio de Janeiro, RJ, 21941-909 - Brazil

# Contents

# 1. Available topologies

## 2D Topologies
Topologies available for constructing bidimensional covalent organic frameworks

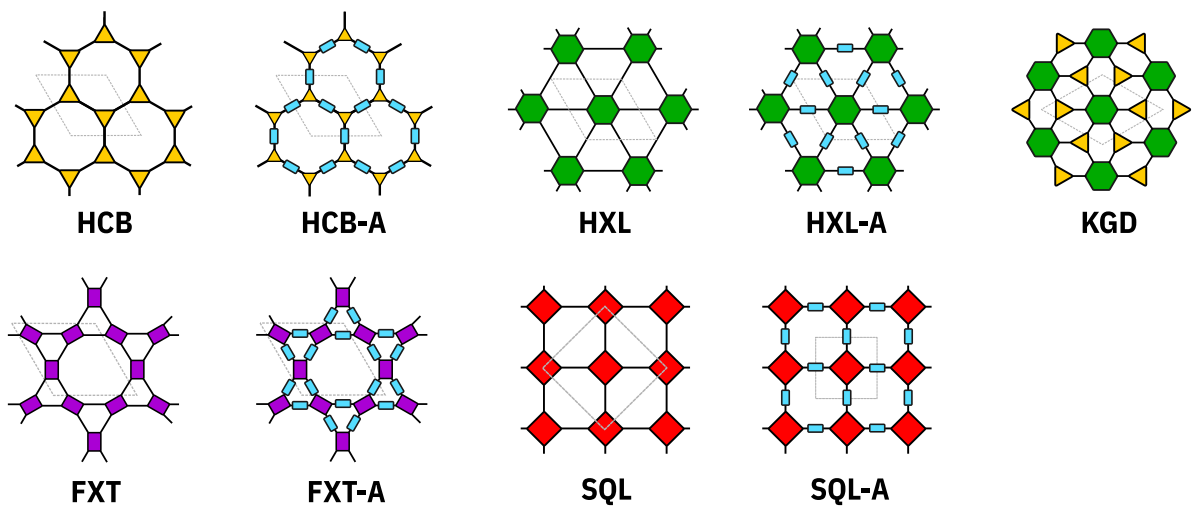

**Figure S1:** Regular and augmented 2D periodic nets currently available on pyCOFBuilder.

All the nets presented on Fig. S1, except the HXL and FXT, can be created by connecting the building blocks using any kind of connection group depicted on Fig. S9 as long as the selected groups are chemically compatible, *i.e.*, can react to form a covalent bond. The HXL and FXT nets only present one kind of node on the unit cell requiring the connection of the selected building block with itself. This kind of connection can only be achieved by coupling reactions, such as Br-Br or Cl-Cl, requiring that both building blocks `A` and `B` be the same molecule. For more details see Section 4.

## 2. Available stacking patterns

# Stacking patterns for hexagonal cells

Different types of stacking patterns for HCB, HCB-A, HXL, HXL-A, KGD, FXT, and FXT-A nets

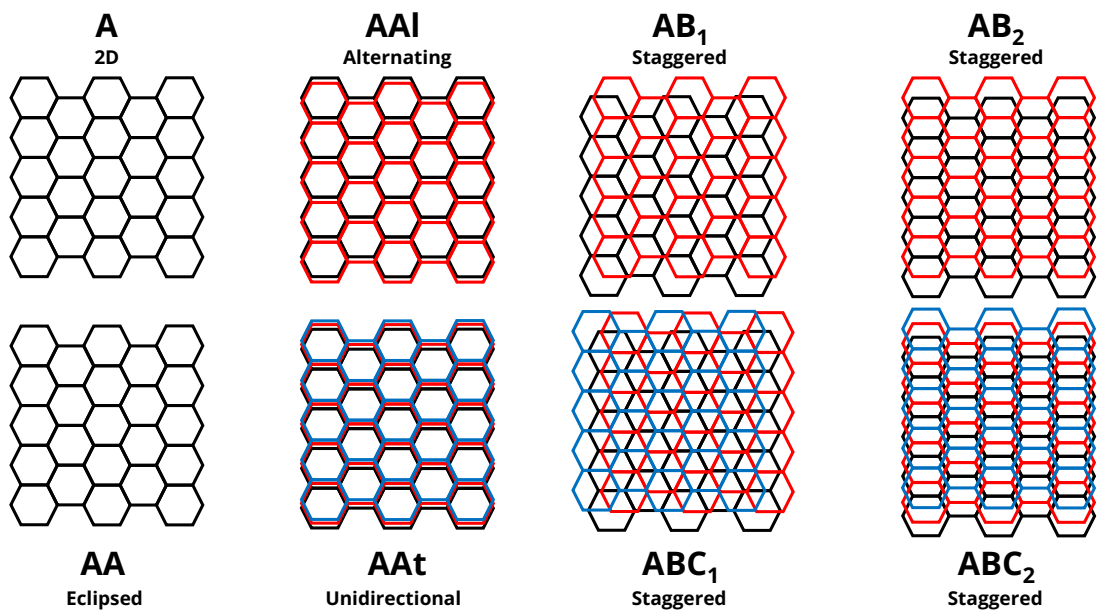

**Figure S2:** Stacking patterns available for 2D hexagonal COFs.

# Stacking patterns for squared cells

Different types of stacking patterns for SQL and SQL-A nets

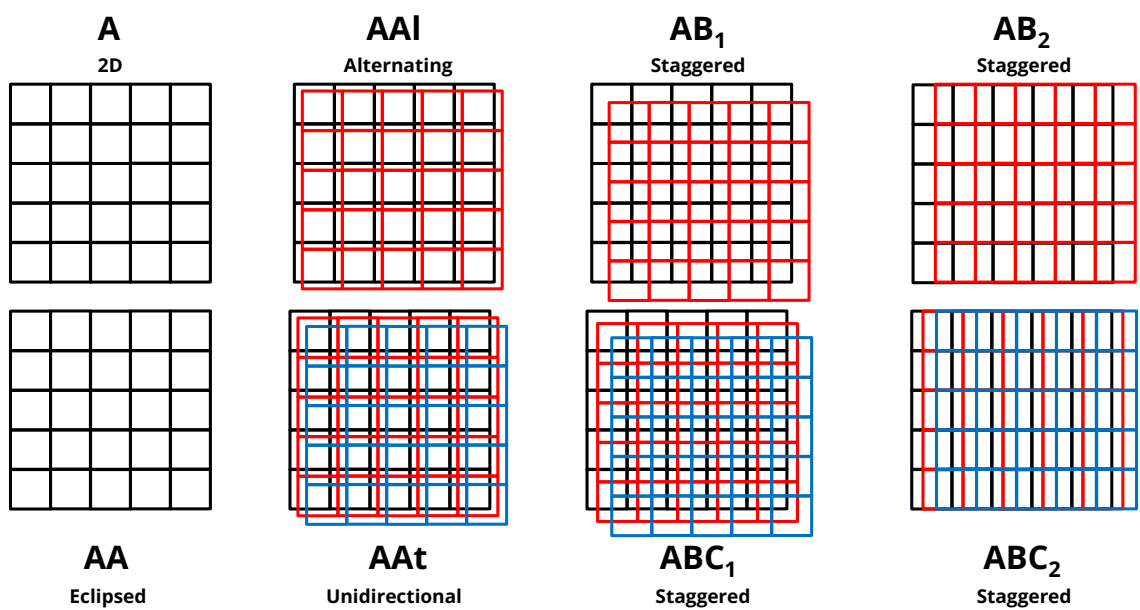

**Figure S3:** Stacking patterns available for 2D square COFs.

Fig. S2 and Fig. S3 present a general overview of the available stacking patterns for hexagonal and square COFs, respectivelly.

## 3. Available organic cores

Fig. S4 and Fig. S5 shows the linear ditopic (L2) building blocks currentli available on pyCOFBuilder. Fig. S6, Fig. S7 and Fig. S8 shows the triangular tritopic, squared tetratopic and hexagonal hexatopic building blocks, respectively.

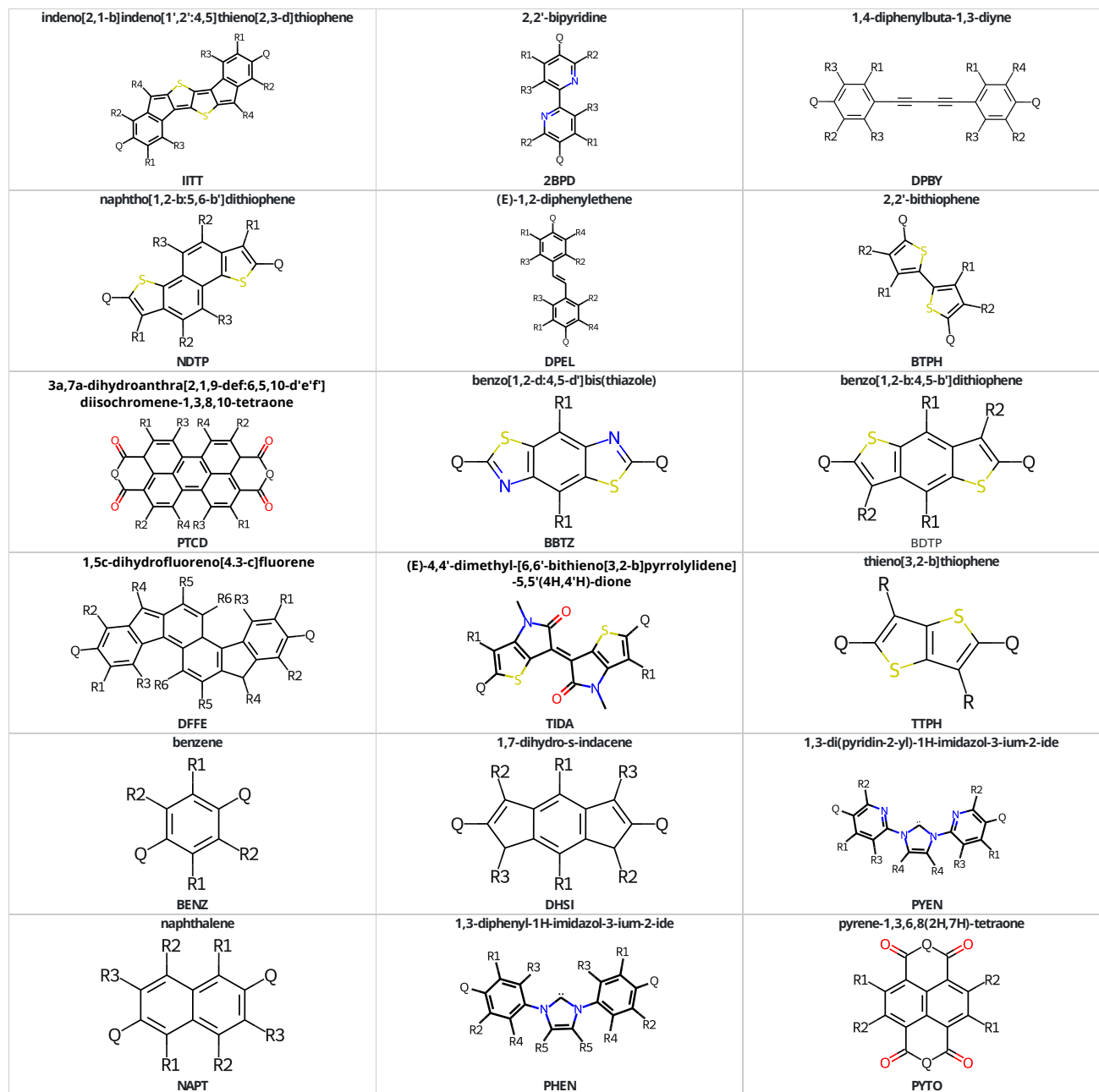

**Figure S4:** Linear ditopic (L2) building blocks part 1.

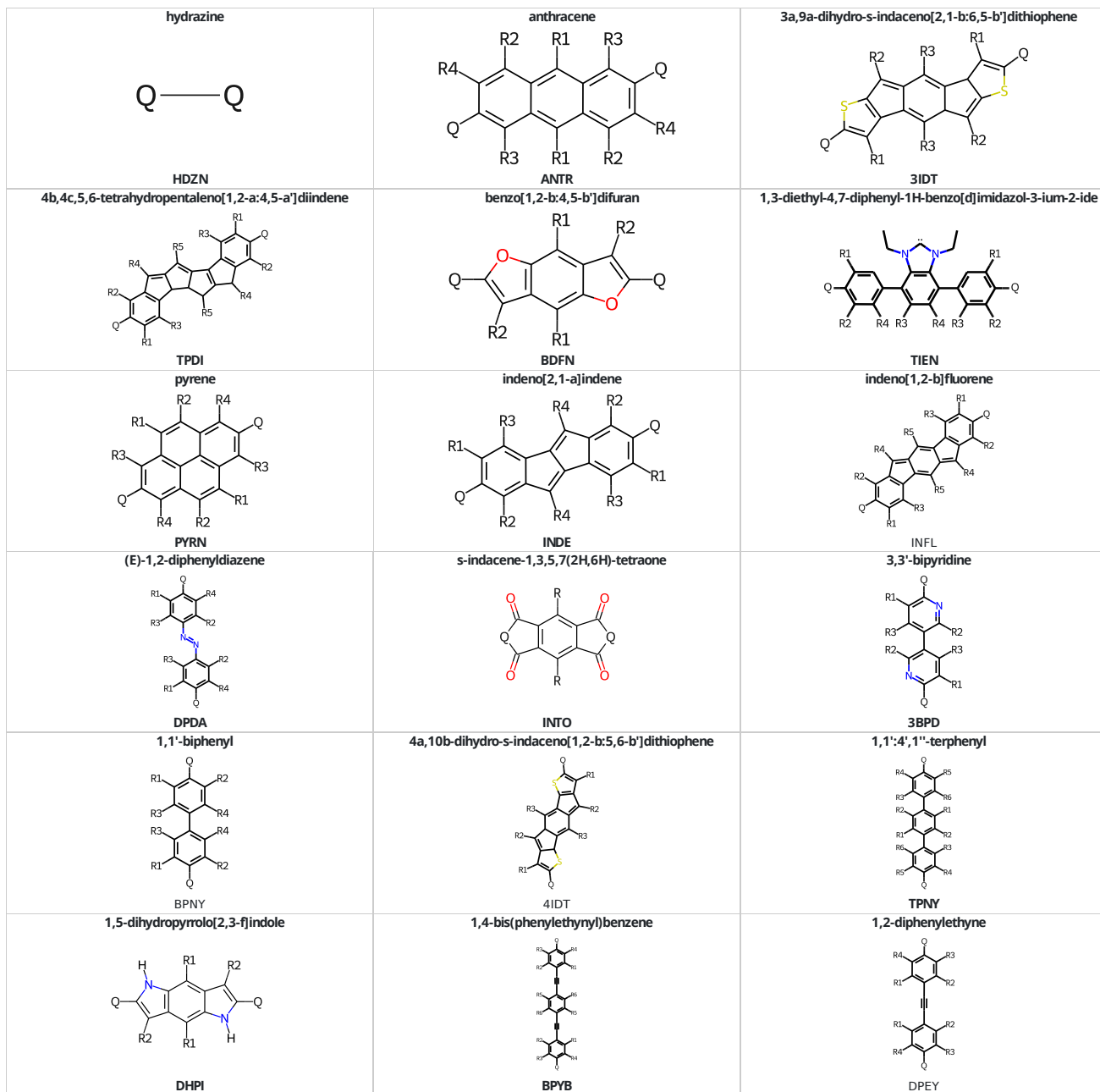

**Figure S5:** Linear ditopic (L2) building blocks part 2.

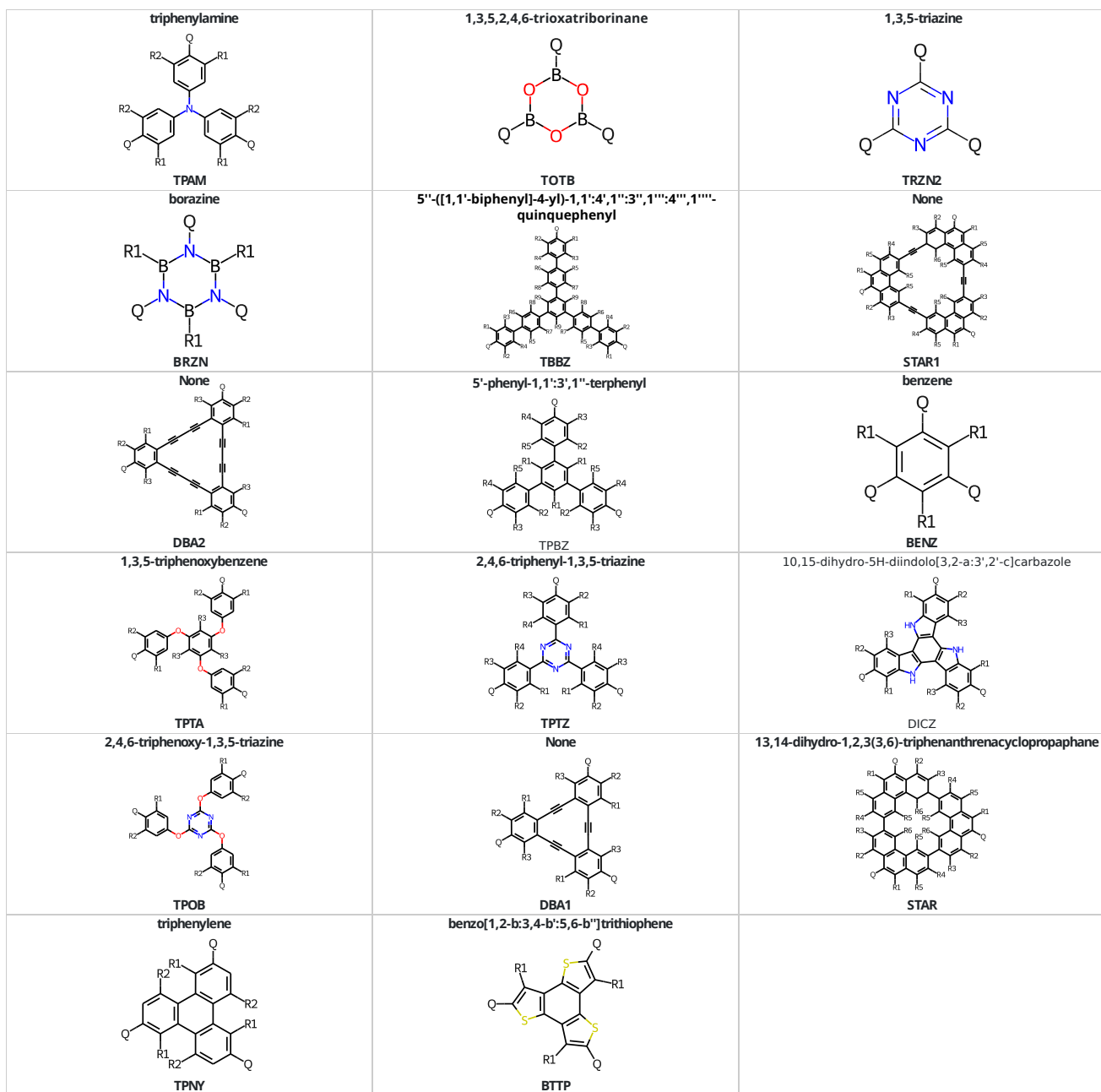

**Figure S6:** Triangular tritopic (T3) building blocks.

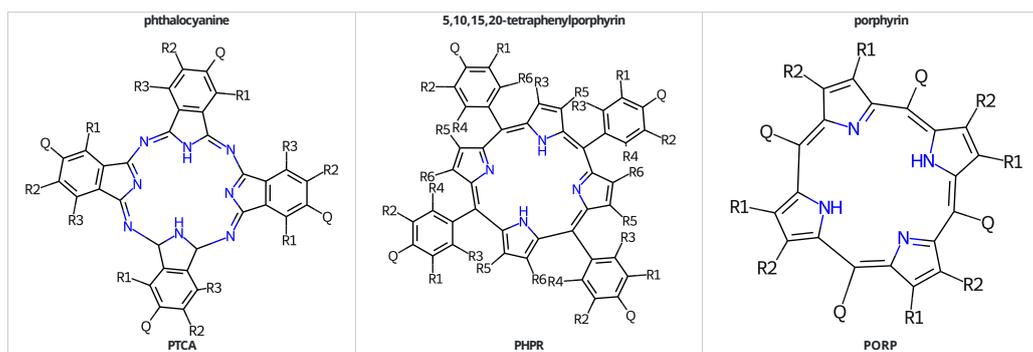

**Figure S7:** Square tetratopic (S4) building blocks.

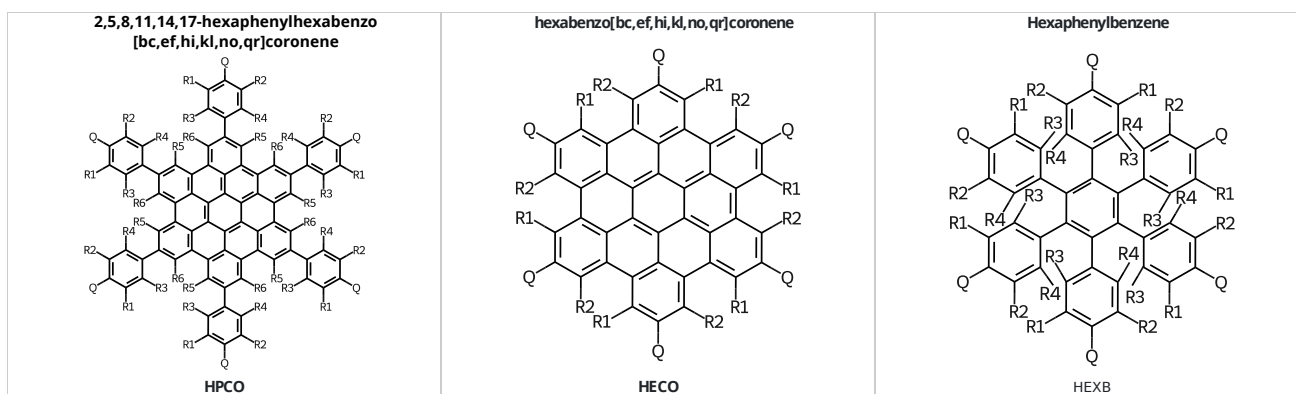

**Figure S8:** Hexagonal hexatopic (H6) building blocks.

## 4. Available connection groups

Fig. S9 shows all the available chemical paths to connect the building blocks into reticular structures implemented on pyCOFBuilder. For the formation of some types of connection groups, such as boronic, borazine, and triazine condensation, it is necessary to decompose the structure using a process similar to the Secondary Building Unit (SBU) approach, where instead of using the correct chemical path to connect the structure an auxiliary path is created decomposing the reticular structure into suitable building units. For more details about this process see Section 6.

# Connection groups

Different types of connection groups available for connection the building blocks into a reticular structure

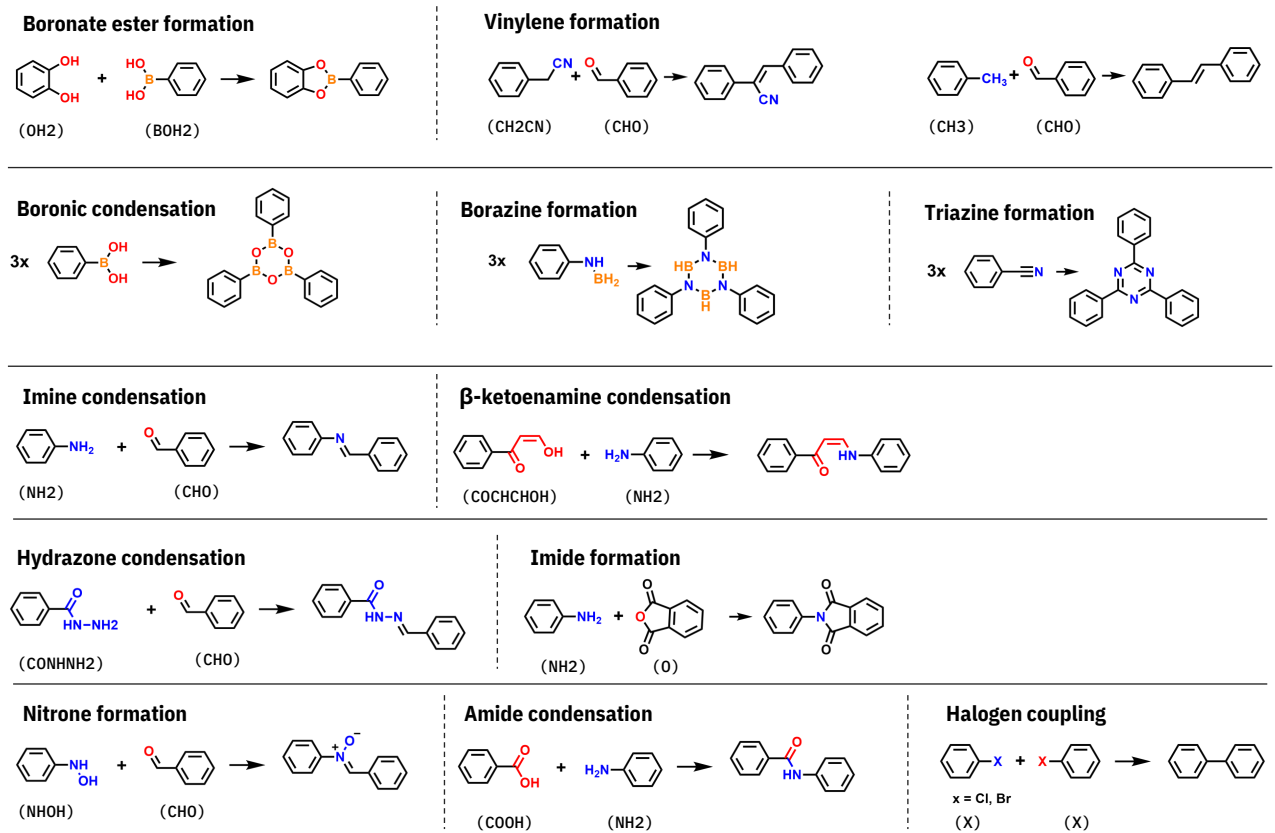

**Figure S9: Available connection groups for bonding the organic cores into reticular structures.** In parenthesis is shown the code for use in pyCOFBuilder. The boronic condensation, borazine formation, and triazine formation cannot be constructed using the connection group representation requiring the use of the SBU approach.

## 5. Available functional groups

# Functional Groups (R$_x$)

Chemical groups available for functionalization of the building blocks

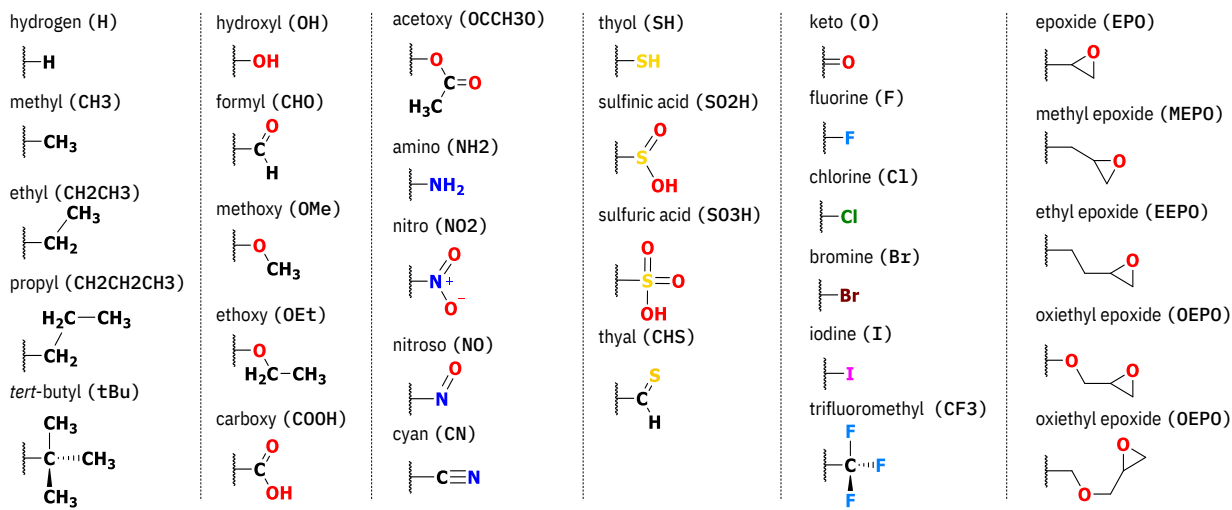

**Figure S10: Available functional groups for functionalizing organic cores.** In parenthesis is shown the code for the use in pyCOFBuilder.

## 6. pyCOFBuilder nomenclature

To establish a useful and comprehensive string-based representation for COF structures, the proposed method must generate new structures based on the reticular approach and allow the representation of already-existent structures. Demonstrative instances are showcased in Table S1 and Table S2, featuring a selection of several well-known 2D and 3D structures along with their corresponding string-based representations. The stacking pattern or interpenetration class was assigned following the reported structure in the publication source.

**Table S1:** pyCOFBuilder string-based representation of some selected 2D COF structures.

| Structure | Original name | pyCOFBuilder Name | Source |
|---|---|---|---|
| 1 | COF-42 | T3_BENZ_CHO_H-L2_BENZ_CONHNH2_OEt-HCB_A-AA | Uribe-Romo et al.[S1] |
| 2 | COF-43 | T3_TPBZ_CHO_H-L2_BENZ_CONHNH2_OEt-HCB_A-AA | Uribe-Romo et al.[S1] |
| 3 | DhaTab | T3_TPBZ_NH2_H-L2_BENZ_CHO_OH-HCB_A-AA | Kandambeth et al.[S2] |
| 4 | TAPB-DMTA | T3_TPBZ_NH2_H-L2_BENZ_CHO_OMe-HCB_A-AA | Xu et al.[S3] |
| 5 | DAB-TFP / TpPa-1 | T3_BENZ_CHO_OH-L2_BENZ_NH2_H-HCB_A-AA | DeBlase et al.[S4] / Kandambeth et al.[S5] |
| 6 | DAAQ-TFP | T3_BENZ_CHO_OH-L2_ANTR_NH2_O-HCB_A-AA | DeBlase et al.[S4] |
| 7 | TpPa-2 | T3_BENZ_CHO_OH-L2_BENZ_NH2_CH3_H-HCB_A-AA | Kandambeth et al.[S5] |
| 8 | COF-JLU2 / RIO-13 | T3_BENZ_CHO_OH-L2_HDZN_NH2-HCB_A-AA | Li et al.[S6] / Maia et al.[S7] |
| 9 | TpBD | T3_BENZ_CHO_OH-L2_BPNY_NH2_H-HCB_A-AA | Han et al.[S8] |
| 10 | TpBD-Me2 | T3_BENZ_CHO_OH-L2_BPNY_NH2_CH3-HCB_A-AA | Han et al.[S8] |
| 11 | TpTab | T3_BENZ_CHO_OH-T3_TPBZ_NH2_H-HCB-AA | Han et al.[S8] |
| 12 | Py-Azine COF | R4_TPPY_CHO_H-L2_HDZN_NH2-KGM_A-AA | Dalapati et al.[S9] |

**Table S2:** pyCOFBuilder string-based representation of some selected 3D COF structures.

| Structure | Original name | pyCOFBuilder Name | Source |
|---|---|---|---|
| 1 | COF-DL229 | T4_BDIA_NH2_H-L2_BENZ_CHO_H-DIA_A-8 | Wang et al.[S10] |
| 2 | COF-300 | T4_TETP_NH2_H_H_H_H_H-L2_BENZ_CHO_H_H-DIA_A-7 | Ma et al.[S11] |
| 3 | COF-303 | T4_TETP_CHO_H_H_H_H_H-L2_BENZ_NH2_H_H-DIA_A-7 | Ma et al.[S11] |
| 4 | LZU-79 | T4_TETP_NH2_H_H_H_H_H-L2_XXXX_CHO_H_H-DIA_A-10 | Ma et al.[S11] |
| 5 | LZU-111 | T4_TETP_NH2_H_H_H_H_H-T4_YYYY_CHO_H_H-LON-3 | Ma et al.[S11] |

Although constructing a consistent representation for the majority of COFs found in the literature is relatively straightforward, certain types of covalent connections can pose a greater challenge due to their impact on the formation of the final structure. For example, COFs formed through boronic condensation between a boronic diacid, such as COF-1[S12], feature only a linear building block, which directly constrains the way the string-based representation is constructed. However, these materials can be perceived as being formed by an HCB_A network, wherein the tritopic building block corresponds to the hexagon formed by O and B atoms (`BRXN`), while the linear ditopic block represents the benzene ring (`BENZ`) present in the original diboronic acid, as illustrated in Figure S11. Consequently, this structure can be represented as `T3_BRXN_BOH2-L2_BENZ_BOH2_H_H_HCB_A-AB`.

The identical approach can be applied to COFs synthesized through the trimerization of dicyanobenzene, resulting in the formation of the triazine-based framework CTF-1.[S13] In this case, the structure can be constructed using the representation `T3_TRZN_CN-L2_BENZ_CN_H_H_HCB_A-AA`.

Another challenge of this nomenclature approach arises when dealing with the representation of multicomponent structures.[S14] These structures are composed of three or more building blocks that can occupy the same reticular site, preventing a straightforward decomposition into one of the existing topological networks. Thus, in the current version of pyCOFBuilder, such structures cannot be constructed.

# Secondary Building Unit (SBU) approach

The self condensation nature on the formation of COF-1 requires a special approach to decompose its structure into a reticular net

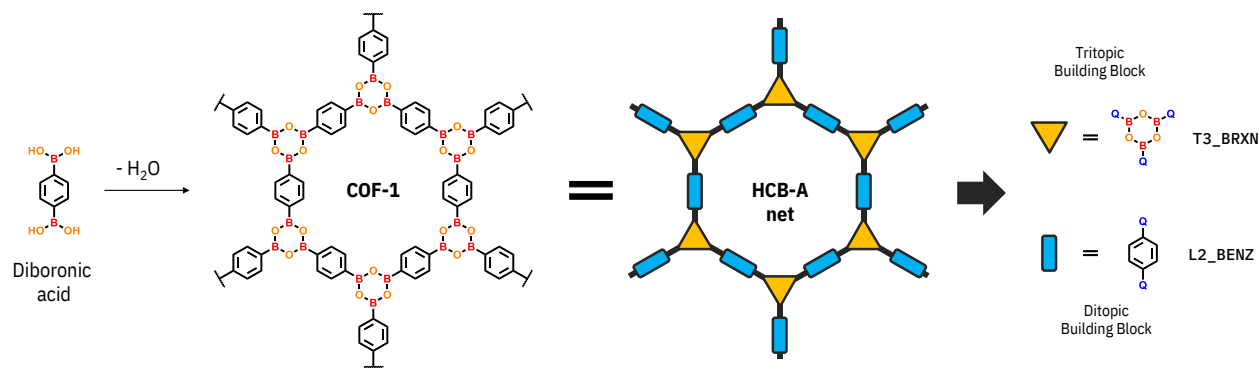

**Figure S11: Illustration of the Secondary Building Unit approach to representing the COF-1 formation from the boronic condensation of the diboronic acid by an HCB-A net.** The self-condensation of the diboronic acid requires a slightly different approach to create the string-based representation, where the portion created by the boronic condensation will be the tritopic triangular block and the aromatic ring portion remaining from the original building block will be the ditopic linear block.

## 7. Experimental data

Table S3: Original name, pyCOFbuilder name, and crystallographic parameters for experimental data of COFs.

| Original name | pyCOFBuilder name | a (Å) | b (Å) | c (Å) | $\alpha$ (°) | $\beta$ (°) | $\gamma$ (°) | Reference |
|---|---|---|---|---|---|---|---|---|
| TAPB-PDA | T3_TPTZ_NH2_H-L2_BENZ_CHO_H-HCB_A-AA | 36.182 | 36.182 | 3.463 | 90 | 90 | 120 | Liu et al.[S15] |
| TAPB-BPDA | T3_TPTZ_NH2_H-L2_BPNY_CHO_H-HCB_A-AA | 44.754 | 44.754 | 3.780 | 90 | 90 | 120 | Liu et al.[S15] |
| TAPB-DMTA | T3_TPTZ_NH2_H-L2_BENZ_CHO_CH3-HCB_A-AA | 35.184 | 35.184 | 3.466 | 90 | 90 | 120 | Liu et al.[S15] |
| TAPB-PDA | T3_TPBZ_NH2_H-L2_BENZ_CHO_H-HCB_A-AA | 37.324 | 37.324 | 3.531 | 90 | 90 | 120 | Matsumoto et al.[S16] |
| TAPB-BPDA | T3_TPBZ_NH2_H-L2_BPNY_CHO_H-HCB_A-AA | 44.738 | 44.738 | 3.531 | 90 | 90 | 120 | Matsumoto et al.[S16] |
| TAPB-TIDA | T3_TPBZ_NH2_H-L2_TIDA_CHO_H-HCB_A-AA | 46.496 | 46.496 | 3.559 | 90 | 90 | 120 | Matsumoto et al.[S16] |
| TPB-DMTA | T3_TPBZ_NH2_H-L2_BENZ_CHO_OMe-HCB_A-AA | 36.205 | 36.205 | 3.884 | 90 | 90 | 120 | Wang et al.[S17] |
| TTA-TTB | T3_TPTZ_NH2_H-T3_TPTZ_CHO_H-HCB-AA | 25.408 | 25.408 | 3.607 | 90 | 90 | 120 | Wang et al.[S17] |
| TPB-DMeTP | T3_TPBZ_NH2_H-L2_BENZ_CHO_CH3-HCB_A-AA | 37.649 | 37.649 | 3.466 | 90 | 90 | 120 | Tao et al.[S18] |
| Tz-TT | T3_TPTZ_NH2_H-L2_TTPH_CHO_H-HCB_A-AA | 38.743 | 38.743 | 3.562 | 90 | 90 | 120 | Ascherl et al.[S19] |
| Star-1 | T3_STAR_OH2_H-L2_BENZ_BOH2_H-HCB_A-AA | 38.520 | 38.520 | 3.380 | 90 | 90 | 120 | Feng et al.[S20] |
| Star-2 | T3_STAR_OH2_H-L2_PYRN_BOH2_H-HCB_A-AA | 46.330 | 46.330 | 3.420 | 90 | 90 | 120 | Feng et al.[S20] |
| Star-3 | T3_STAR_OH2_H-L2_BPNY_BOH2_H-HCB_A-AA | 46.830 | 46.830 | 3.410 | 90 | 90 | 120 | Feng et al.[S20] |
| DBA-COF 1 | T3_DBA1_OH2_H-L2_BENZ_BOH2_H-HCB_A-AA | 33.929 | 33.929 | 3.400 | 90 | 90 | 120 | Baldwin et al.[S21] |
| DBA-COF 2 | T3_DBA2_OH2_H-L2_BENZ_BOH2_H-HCB_A-AA | 37.889 | 37.889 | 3.400 | 90 | 90 | 120 | Baldwin et al.[S21] |
| Tp-PA-1 | T3_BENZ_CHO_OH-L2_BENZ_NH2_H-HCB_A-AA | 22.194 | 22.194 | 3.454 | 90 | 90 | 120 | Han et al.[S8] |
| TpBD | T3_BENZ_CHO_OH-L2_BPNY_NH2_H-HCB_A-AA | 29.800 | 29.800 | 3.418 | 90 | 90 | 120 | Han et al.[S8] |
| TpBD-Me2 | T3_BENZ_CHO_OH-L2_BPNY_NH2_CH3-HCB_A-AA | 29.581 | 29.581 | 3.461 | 90 | 90 | 120 | Han et al.[S8] |
| TpTab | T3_BENZ_CHO_OH-T3_TPBZ_NH2_H-HCB-AA | 18.344 | 18.344 | 3.459 | 90 | 90 | 120 | Han et al.[S8] |
| COF-5 | T3_TPNY_OH2_H-L2_BENZ_BOH2_H-HCB_A-AA | 29.700 | 29.700 | 3.460 | 90 | 90 | 120 | Cote et al.[S12] |
| COF-6 | T3_BENZ_BOH2_H-T3_TPNY_OH2_H-HCB-AA | 14.974 | 14.974 | 3.398 | 90 | 90 | 120 | Cote et al.[S22] |
| COF-8 | T3_TPBZ_BOH2_H-T3_TPNY_OH2_H-HCB-AA | 22.013 | 22.013 | 3.630 | 90 | 90 | 120 | Cote et al.[S22] |
| COF-10 | T3_TPNY_OH2_H-L2_BPNY_BOH2_H-HCB_A-AA | 36.028 | 36.028 | 3.526 | 90 | 90 | 120 | Cote et al.[S22] |
| TAPB-DMTA | T3_TPBZ_NH2_H-L2_BENZ_CHO_OMe-HCB_A-AA | 36.459 | 37.154 | 3.523 | 90 | 90 | 120 | Xu et al.[S3] |
| TAPB-PTCDA | T3_TPBZ_NH2_H-L2_PTCD_O_H-HCB_A-AA | 43.600 | 43.600 | 3.705 | 90 | 90 | 120 | Maschita et al.[S23] |
| TAPB-PMDA | T3_TPBZ_NH2_H-L2_INTO_O_H-HCB_A-AA | 35.700 | 35.700 | 3.760 | 90 | 90 | 120 | Maschita et al.[S23] |
| TAPA-PMDA | T3_TPAM_NH2_H-L2_INTO_O_H-HCB_A-AA | 31.100 | 31.100 | 4.018 | 90 | 90 | 120 | Maschita et al.[S23] |
| TT-PMDA | T3_TPTZ_NH2_H-L2_INTO_O_H-HCB_A-AA | 35.200 | 35.200 | 3.700 | 90 | 90 | 120 | Maschita et al.[S23] |
| DAAQ-TFP | T3_BENZ_CHO_OH-L2_ANTR_NH2_O-HCB_A-AA | 29.833 | 29.833 | 3.621 | 90 | 90 | 120 | DeBlase et al.[S4] |
| TAPB-DMPDA | T3_TPBZ_NH2_H-L2_BENZ_CHO_OMe_H-HCB_A-AA | 36.358 | 36.358 | 3.512 | 90 | 90 | 120 | Natraj et al.[S24] |
| RC-COF-1 | T3_BENZ_CHO_OH-L2_BENZ_NH2_H-HCB_A-AA | 22.040 | 22.040 | 3.490 | 90 | 90 | 120 | Zhang et al.[S25] |
| RC-COF-2 | T3_BENZ_CHO_OH-L2_BPNY_NH2_CH3-HCB_A-AA | 29.290 | 29.290 | 3.530 | 90 | 90 | 120 | Zhang et al.[S25] |
| RC-COF-3 | T3_BENZ_CHO_OH-L2_BPNY_NH2_H-HCB_A-AA | 29.050 | 29.050 | 3.580 | 90 | 90 | 120 | Zhang et al.[S25] |
| RC-COF-4 | T3_BENZ_CHO_OH-L2_DPEY_NH2_H-HCB_A-AA | 32.100 | 32.100 | 3.160 | 90 | 90 | 120 | Zhang et al.[S25] |
| TTI-COF | T3_TPTZ_CHO_H-T3_TPTZ_NH2_H-HCB-AA | 26.060 | 26.060 | 3.674 | 90 | 90 | 120 | Zhang et al.[S25] |
| MeDBP-TFB | T3_BENZ_CHO_H-L2_INDE_NH2_CH3-HCB_A-AA | 32.470 | 32.470 | 3.840 | 90 | 90 | 120 | Sprachmann et al.[S26] |
| PhDBP-TFB | T3_BENZ_CHO_OH-L2_INDE_NH2_Ph-HCB_A-AA | 33.100 | 33.100 | 3.400 | 90 | 90 | 120 | Sprachmann et al.[S26] |
| TPB-imine-COF | T3_TPBZ_NH2_H-L2_BPNY_CHO_NO2-HCB_A-AA | 42.940 | 42.940 | 3.440 | 90 | 90 | 120 | Sprachmann et al.[S26] |

## 8. GCMC simulations of CO$_2$ adsorption

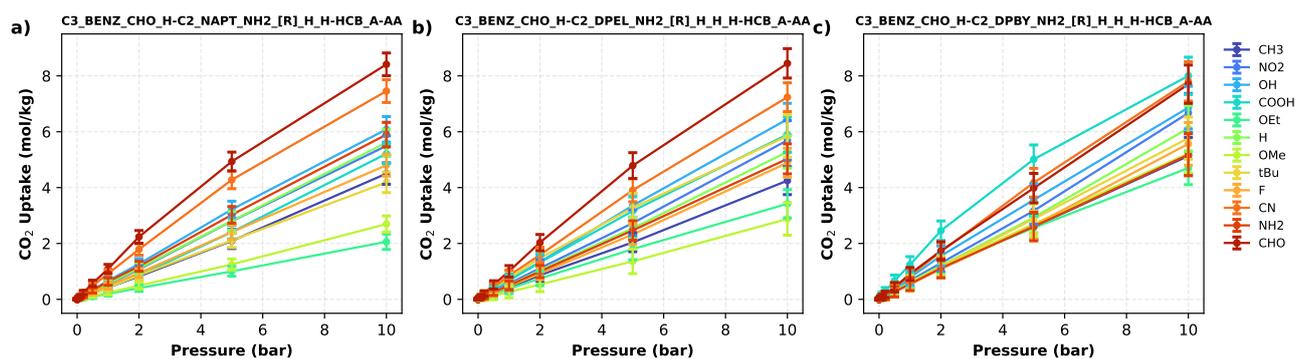

Figure S12: Impact of functional groups on the CO$_2$ adsorption of three different COFs.

The Fig. S12 shows the impact of different functional groups on the CO$_2$ adsorption of three different COFs built from the connection of a C3_BENZ_CHO_H building block with a C2_NAPT_NH2_[R], C2_DPEL_NH2_[R], and C2_DPBY_NH2_[R], where [R] represents the functional groups.

Fig. S13 and Fig. S14 show the parity plot showing the xGBoost predicted versus the reference value for the working capacity and the selectivity.

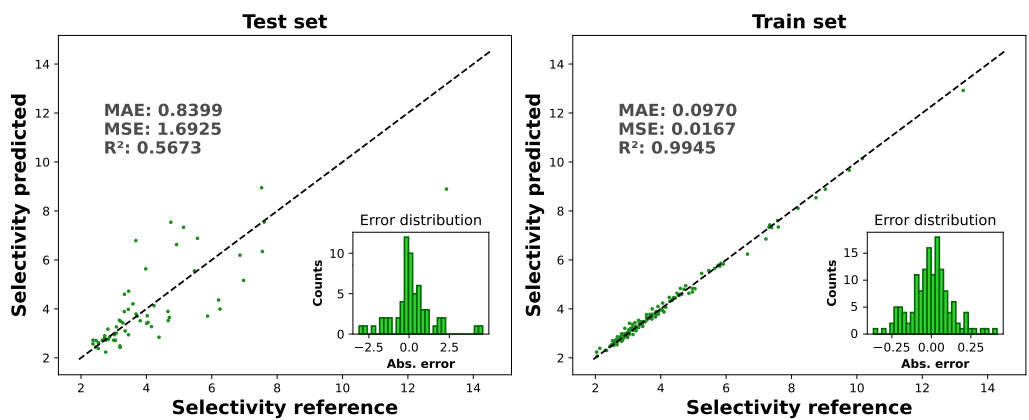

Figure S13: Parity plot of the predicted values obtained from the xGBoost model for predicting the selectivity. Test set (left) and train set (right).

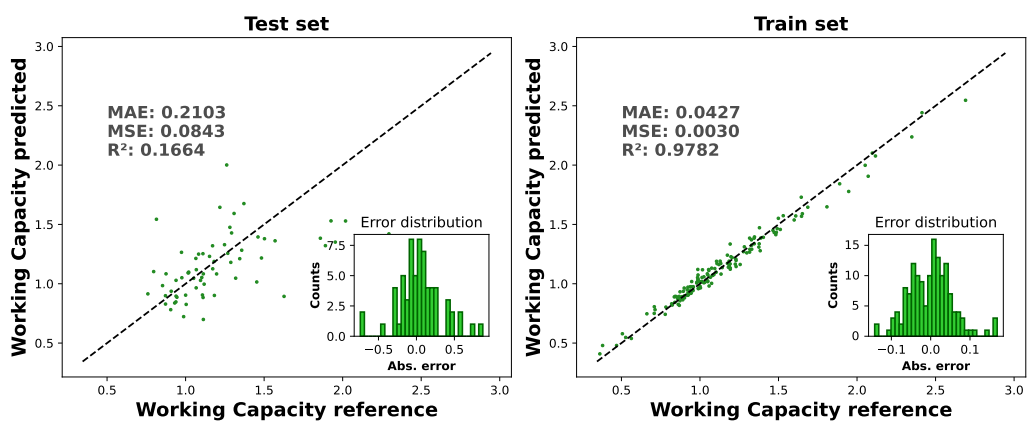

Figure S14: Parity plot of the predicted values obtained from the xGBoost model for predicting the working capacity. Test set (left) and train set (right).